\documentclass[reprint,showpacs,preprintnumbers,amsmath,amssymb,prc,floatfix]{revtex4-1}
\usepackage{float}
\usepackage{color}
\usepackage{graphicx}
\usepackage{dcolumn}
\usepackage{bm}
\usepackage{xcolor}
\usepackage{comment}
\usepackage[colorlinks,citecolor=blue,linkcolor=red,anchorcolor=blue,filecolor=blue,urlcolor=blue]{hyperref}
\usepackage{multirow}
\usepackage{threeparttable} 
\usepackage{adjustbox} 
\usepackage{color}
\usepackage{longtable}
\usepackage{csquotes}{}

\begin{document}


\title{
Influence of effective interactions and nuclear densities on the dynamics of heavy-ion fusion} 

\author{Shilpa Rana$^1$}
\email{shilparana1404@gmail.com}
\author{Raj Kumar$^1$}
\email{rajkumar@thapar.edu}
\author{M. Bhuyan$^2$}
\email{mrutunjaya.b@iopb.res.in}

\bigskip
\affiliation{$^1$Department of Physics and Materials Science, Thapar Institute of Engineering and Technology, Patiala, Punjab 147004, India}
\affiliation{$^2$Institute of Physics, Sachivalaya Marg, Bhubaneswar 751005, Odisha, India} 
\bigskip

\begin{abstract}
\noindent
The study aims to explore the mechanism of heavy-ion fusion using various effective nucleon-nucleon (NN) interactions and nuclear density distributions. The nuclear potentials are obtained by folding the relativistic effective NN interaction (R3Y) potential with the densities derived from the relativistic mean-field (RMF) approach. Here, we have considered the RMF formalism with a non-linear NL3$^*$ parameter set and the relativistic-Hartree-Bogoliubov (RHB) approach with the DDME2 parameter set to obtain the medium-independent R3Y and density-dependent R3Y (DDR3Y) NN potentials, respectively. The results of these relativistic effective NN interactions are compared with the well-known Reid and Paris M3Y NN interactions and their density-dependent versions. Nuclear potentials are also used to calculate the fusion barrier characteristics and cross-section within the $\ell-$summed Wong model for $^{16}$O+$^{144}$Sm, $^{48}$Ca+$^{208}$Pb, $^{16}$O+$^{154}$Sm and $^{48}$Ca+$^{238}$U reactions. The relativistic R3Y and DDR3Y NN potentials are noted to provide a higher cross-section than both the Reid and Paris versions of the non-relativistic M3Y (Michigan 3 Yukawa) NN potentials. Moreover, the inclusion of in-medium effects in both relativistic and non-relativistic effective NN interactions is noted to provide relatively repulsive nuclear potentials, which lead to higher fusion barriers and hindered cross-sections. The impact of nuclear shape degrees of freedom in terms of nuclear deformation is also included in the description of nuclear fusion for $^{16}$O+$^{154}$Sm and $^{48}$Ca+$^{238}$U reactions involving deformed target nuclei. Furthermore, the results of RMF-NL3$^*$ densities are also compared with those obtained using densities from the two-parameter Fermi (2pF) formula for the representative case of $^{16}$O+$^{154}$Sm reaction, and the RMF-NL3$^*$ densities are noted to provide a higher cross-section for the considered reaction. The quantitative analysis of cross sections calculated using different nuclear potentials with the available experimental data shows that the R3Y NN potential and nuclear densities calculated within the RMF formalism for the non-linear NL3$^*$ parameter set provide a reasonable agreement with the experimental data for all the considered reactions containing spherical and deformed target nuclei.
\end{abstract}
\maketitle

\section{INTRODUCTION}
\label{intro}
The study of heavy-ion fusion reactions is important to understand various phenomena that range from the evolution of massive stars to the extension of the periodic table through the synthesis of various exotic nuclei. These exotic nuclei are crucial for exploring the behaviour of nuclear matter under extreme conditions, which provides insights into the intricate correlation between nuclear structure and reaction dynamics. Consequently, the study of heavy-ion reactions have been the central topic of research for both experimental and theoretical nuclear physicists \cite{canto20,mont17,toub17,das98,raj20,jiang21,stef10,back14,wakhle18,bala95,hagino03}. Theoretical investigations of heavy-ion fusion generally begin with the calculation of the total interaction potential formed between the colliding projectile and the target nuclei \cite{mont17,toub17,canto20,das98,raj20,back14}. However, the underlying uncertainties in the exact form of the strong nuclear forces pose complexities in the description of attractive nuclear potential formed between two fusing heavy-ions. Moreover, nuclear fusion is affected by numerous factors such as degrees of freedom of nuclear shape, corrections to the nuclear shell, pairing energy, incompressibility of nuclear matter, energy dependence of the interaction potential, mass, charge and isospin asymmetry of the entrance channel \cite{mont17,toub17,canto20,das98,raj20}. In the literature, some studies have been carried out to probe the role of one or another of these structural properties on the dynamics of heavy-ion reaction employing nuclear interaction potential obtained within different phenomenological, semi-microscopic and microscopic approaches \cite{raj11,deep13,rajk11,deep14,back14,bloc77,bloc81,cheng19,bass73,bass74,bass77,vaut72,raj20,chamon02,chamon21}. In these theoretical models, stepwise refinements are meticulously carried out to thoroughly investigate the influence of various structural effects on the nuclear reaction mechanism \cite{raj11,deep13,rajk11,deep14,back14,bloc77,bloc81,cheng19,bass73,bass74,bass77,vaut72,raj20,chamon02,chamon21}.

The double folding model \cite{satc79} is one of the well-known approaches to obtain the nuclear interaction potential formed between the two colliding heavy-ions. The densities of the interacting nuclei and an effective nucleon-nucleon (NN) interaction are prerequisites to calculating the nuclear potential within the double-folding approach. In Refs. \cite{sing12,sahu14,lahi16}, the relativistic effective NN interaction named as R3Y potential was derived from the widely adopted relativistic mean-field (RMF) formalism. It was used along with the RMF nuclear densities to study the cluster radioactivity. Later, the R3Y NN potential, along with the RMF densities, was adopted for the first time to explore the fusion dynamics of a few Ni-based reactions \cite{bhuy18}. In \cite{bhuy20}, various heavy-ion reactions leading to the synthesis of heavy and superheavy nuclei were considered to assess the application of this new theoretical approach, i.e., the relativistic mean-field for the study of fusion. The results of the relativistic R3Y NN potential are usually compared with the Reid version of the widely adopted Michigan 3 Yukawa (M3Y) effective NN potential \cite{bert77}. In the recent work of Refs. \cite{bhuy18,bhuy20}, the spherically symmetric densities of the interacting nuclei were used as a simplifying assumption, as it was the first step towards implementing a new theoretical formalism. However, in Ref. \cite{chus22}, concerns were raised about the use of spherical densities for the target nuclei in \cite{bhuy20}.

As mentioned above, the nuclear fusion of two heavy-ions is a complex phenomenon which is affected by different structural properties of interacting nuclei and refinements in the theoretical models are necessary to account for these effects. Following this, the in-medium effects were also accounted for in the description of relativistic R3Y NN potential in terms of density-dependent nuclear meson couplings \cite{bhuy22,rana22}. Out of various structural properties, the influence of nuclear deformations on fusion mechanisms have been discussed quite extensively in the literature \cite{mont17,toub17,canto20,das98,raj20,back14}. Following this, the nuclear shape degrees of freedom are also included in the calculation of microscopic nuclear potential within the RMF formalism \cite{rana23}. The present study aims to conduct a systematic and quantitative analysis of the nuclear fusion mechanism using the double-folding approach supplemented with RMF formalism. First, we aim to explore the effect of different effective NN interactions on the dynamics of nuclear fusion. For this, we have considered two versions of relativistic effective NN interaction, i.e. the R3Y  and density-dependent R3Y (DDR3Y)  NN potentials. The R3Y NN potential is obtained for the non-linear NL3$^*$ parameter set \cite{lala09}, and the well-known DDME2 set \cite{lala05} is used to calculate the medium-dependent DDR3Y NN potential. The results of the relativistic NN interactions are also compared with the density-independent and density-dependent versions of well-known Reid and Paris M3Y NN interactions \cite{satc79,kobo82,khoa94,bert77,anan83}. The densities of interacting nuclei are obtained within the RMF approach with a non-linear NL3$^*$ parameter set and the relativistic-Hartree-Bogoliubov (RHB) with a density-dependent DDME2 parameter set. The results of the RMF densities will also be compared with those obtained using densities from the two-parameter Fermi (2pF) formula. Further, the role of nuclear shape degrees of freedom on the nuclear fusion mechanism will be explored by including the target deformation in the calculation of RMF densities. Here, we have considered $^{16}$O+$^{144}$Sm and $^{48}$Ca+$^{208}$Pb reactions containing the spherical reactions partners as well as $^{16}$O+$^{154}$Sm and $^{48}$Ca+$^{238}$U reactions involving deformed targets. The fusion and/or capture cross-section is obtained within the $\ell-$summed Wong model \cite{kuma09,wong73}, and results are also compared with the available experimental data \cite{morton94,prok08,leigh95,nishi12}.

The paper is organized as follows: The description of adopted theoretical formalism is provided in Sec. \ref{theory}. The results are discussed in Sec. \ref{results}, and Sec. \ref{summary} contains the summary and conclusions of the present study.  

\section{THEORETICAL FORMALISM}
\label{theory}
In the process of heavy-ion collision, the interplay between the attractive and repulsive components of total interaction potential results in the formation of a fusion barrier. A suitable description of the total interaction potential is essential to study the characteristics of the fusion barrier. The interaction potential formed between a spherically symmetric projectile fusing with a quadrupole-deformed target can be written as,    
\begin{eqnarray}
V_T(R,\beta_2,\theta_2) \!\!= \!\!V_C(R,\beta_2,\theta_2)\!+\!V_n(R,\beta_2,\theta_2)\!+\!\frac{\hbar^2\ell(\ell+1)}{2\mu R^2}. 
\nonumber \\ 
\label{vtot}
\end{eqnarray}
Here, $R$ is the inter-nuclear separation, and $\beta_2$ denotes the quadrupole deformation of the target nucleus. $\theta_2$ is the orientation angle between the symmetry axis of the deformed target nucleus and $R$. $\mu$ symbolizes the reduced mass of the target-projectile system and $V_C(R,\beta_2,\theta_2)$ is the deformation and orientation dependent repulsive Coulomb potential \cite{wong73}. The term $V_n(R,\beta_2,\theta_2)$ denotes the attractive nuclear potential and is calculated within the double folding approach \cite{satc79}. In the folding method, nuclear potentials are obtained by folding an effective NN interaction with nuclear density distributions, allowing for a realistic description of optical potentials and reaction dynamics  \cite{satc79,kobo82,khoa94}, i.e.,
\begin{eqnarray}
V_{n}(\vec{R},\beta_2,\theta_2) \!\!\! &=& \!\!\!\! \int \!\! \rho_{p}(\vec{r}_p)\rho_{t}(\vec{r}_t(\beta_2,\theta_2))\nonumber \\
&& V_{eff}
\left( \boldsymbol{\rho},r{\equiv}|\vec{r}_p-\vec{r}_t +\vec{R}|  \right) d^{3}r_pd^{3}r_t.
\label{fold}
\end{eqnarray}
Here, $\rho_p(\vec{r}_p)$ and $\rho_t(\vec{r}_t(\beta_2,\theta_2))$ denote the total densities (sum of the proton and neutron densities) of the projectile and target nuclei, respectively. $V_{eff}$ is the effective nucleon-nucleon (NN) interaction. These quantities are the basic requirements to calculate the nuclear potential within the double-folding model. Bare NN potentials are fundamental interactions derived from NN scattering data, describing the force between two free nucleons without medium effects. However, due to the complexities of the nuclear many-body system, these bare NN potentials are not directly applicable in nuclear reaction studies \cite{satc79}. Instead, effective NN interactions are employed. These are modified versions of the bare NN interactions that account for medium effects such as Pauli blocking and density dependence, thereby making them suitable for reaction calculations. The Michigan 3 Yukawa (M3Y) interaction is one such effective NN potential widely used in nuclear reaction studies \cite{satc79,kobo82,khoa94,bert77,anan83}. The M3Y interaction was introduced as an effective force derived from realistic NN potentials, primarily for use in optical model and folding potential calculations \cite{bert77}. It is important to note that this M3Y potential is distinct from the phenomenological approach presented in Ref. \cite{myers66}, which is employed in a macroscopic-microscopic framework to describe nuclear masses and deformations via a type of density functional.

In the present study, the inputs of the double folding model, i.e., nuclear densities and effective NN interaction, are obtained within the relativistic mean-field (RMF) formalism. The RMF formalism has been extensively utilized to investigate the properties of finite nuclei, lying both near and far from the regions of $\beta$-stability \cite{vret05,meng16,ring96,lala09,lala97,rein86,sing12,sahu14,lahi16,meng06,dutra14,afan05}. In the RMF approach, the inter-nucleon interaction is considered to be mediated through the exchange of photons and mesons. A phenomenological Lagrangian density, which describes the nuclear-meson many-body interactions, can be formulated as follows \cite{vret05,meng16,ring96,lala09,lala97,rein86,sing12,sahu14,lahi16,meng06,dutra14,afan05}. 
 \begin{eqnarray}
{\cal L}&=&\overline{\psi}\{i\gamma^{\mu}\partial_{\mu}-M\}\psi +{\frac12}\partial^{\mu}\sigma
\partial_{\mu}\sigma \nonumber \\
&& -{\frac12}m_{\sigma}^{2}\sigma^{2}-{\frac13}g_{2}\sigma^{3} -{\frac14}g_{3}\sigma^{4}
-g_{\sigma}\overline{\psi}\psi\sigma \nonumber \\
&& -{\frac14}\Omega^{\mu\nu}\Omega_{\mu\nu}+{\frac12}m_{\omega}^{2}\omega^{\mu}\omega_{\mu}
-g_{w}\overline\psi\gamma^{\mu}\psi\omega_{\mu} \nonumber \\
&&-{\frac14}\vec{B}^{\mu\nu}.\vec{B}_{\mu\nu}+\frac{1}{2}m_{\rho}^2
\vec{\rho}^{\mu}.\vec{\rho}_{\mu} -g_{\rho}\overline{\psi}\gamma^{\mu}
\vec{\tau}\psi\cdot\vec{\rho}^{\mu}\nonumber \\
&&-{\frac14}F^{\mu\nu}F_{\mu\nu}-e\overline{\psi} \gamma^{\mu}
\frac{\left(1-\tau_{3}\right)}{2}\psi A_{\mu}.
\label{lag}
\end{eqnarray}
Here, $\psi$ represents a Dirac nucleon with mass $M$. The terms $m_\sigma$, $m_\omega$, and $m_\rho$ denote the masses of the $\sigma$, $\omega$, and $\rho$ mesons, respectively, while $g_\sigma$, $g_\omega$, and $g_\rho$ are the corresponding nucleon-meson coupling constants. The constants $g_2$ and $g_3$ represent the non-linear self-interactions of the isoscalar scalar $\sigma$ mesons. The meson masses, nucleon-meson coupling constants, and meson self-coupling constants are key parameters in the RMF formalism, typically adjusted to match experimental data for ground state observables of selected closed-shell nuclei. Numerous RMF parameter sets are available in the literature \cite{dutra14}. Here, we have used the non-linear NL3$^*$ parameter set \cite{lala09}, which is a revised version of the widely adopted NL3 set \cite{lala97}. The non-linear couplings ($g_2$ and $g_3$) in the Eq. \eqref{lag} effectively account for the in-medium effects \cite{boguta77}. Another method to introduce the in-medium effects in the inter-nucleon interaction is density-dependent parametrization within the Relativistic-Hartree-Bogoliubov (RHB) approach, where the nucleon-meson couplings are medium-dependent  and are defined as  \cite{niks02,lala05,type99,fuchs95,hofm01}, 
\begin{eqnarray}
g_i( {\rho})=g_i( {\rho}_{sat})f_i(x)|_{i=\sigma,\omega},
\label{dd1}
\end{eqnarray}
where
\begin{eqnarray}
f_i(x)=a_i\frac{1+b_i(x+d_i)^2}{1+c_i(x+d_i)^2}
\label{dd2}
\end{eqnarray}
and
\begin{eqnarray}
g_\rho( {\rho})=g_\rho( {\rho}_{sat})exp[-a_\rho(x-1)].
\label{dd3}
\end{eqnarray}
Here, $x= {\rho} /  {\rho}_{sat}$, with $ {\rho}_{sat}$ denoting the saturation density of symmetric nuclear matter. The five constraints- $f_i(1)=1$, $f''_i(0)=0$, and $f''_\sigma(1)=f''_\omega(1)$ reduce the number of independent parameters in Eq. (\ref{dd2}) from eight to three. The independent parameters, i.e., the meson mass and coupling parameters of the RHB formalism are also calibrated to match the ground state properties of finite nuclei as well as the characteristics of symmetric and asymmetric nuclear matter. In the present analysis, we have adopted the well-known DDME2 parameter sets \cite{lala05} to study the heavy-ion fusion mechanism. The field equations for the Dirac nucleons and inter-mediating mesons are obtained from the Lagrangian density by using the Euler-Lagrange equations under the mean-field approximation. The details of these RMF equations can be found in Refs. \cite{vret05,meng16,ring96,lala09,lala97,rein86,sing12,sahu14,lahi16,meng06,dutra14,afan05}. The relativistic effective NN interaction (R3Y) potential is derived by solving the RMF equations for mesons in the limit of one-pion exchange. This R3Y  NN potential is written in terms of meson-nucleon couplings as,
\begin{eqnarray}
V_{eff}^{R3Y}(r,{\rho}_p, {\rho}_t)&=&\sum_{i=\omega,\rho}\frac{{g_{i}( {\rho}_p) g_{i}( {\rho}_t)}}{4{\pi}}\frac{e^{-m_{i}r}}{r} \nonumber \\
& - &\frac{{g_{\sigma}( {\rho}_p) g_{\sigma}( {\rho}_t)}}{4{\pi}} 
\frac{e^{-m_{\sigma}r}}{r} +\frac{g_{2}^{2}}{4{\pi}} r e^{-2m_{\sigma}r}\nonumber \\
&& +\frac{g_{3}^{2}}{4{\pi}}\frac{e^{-3m_{\sigma}r}}{r} + J_{00}\delta(r). 
\label{r3y}
\end{eqnarray}
It is to be noted here that we have followed the methodology prescribed in Refs. \cite{sch51,sch51a} for the solution of the non-linear equations for  $\sigma$-field containing $g_2$ and $g_3$. The term $J_{00}(E)\delta(r)$ in Eq. \eqref{r3y} represents a pseudo-potential that accounts for the long-range one-pion exchange potential (OPEP). For the case of R3Y NN potential obtained for non-linear NL3$^*$ parameter, the nucleon-meson couplings ($g_\sigma$, $g_\omega$ and $g_\rho$) in Eq. \eqref{r3y} are independent of the density. However,  for the density-dependent R3Y (DDR3Y) NN potential obtained for the DDME2 parameter set, the $g_\sigma$, $g_\omega$ and $g_\rho$ are density-dependent as given in Eqs. (\ref{dd1}-\ref{dd3}). Additionally,  the non-linear self-interaction constants ($g_2$ and $g_3$) are zero for DDR3Y NN potential. Here, we have used both the projectile and target densities to include the medium dependence in the relativistic DDR3Y NN potential within the RHB approach and the terms $[|g_{i}(\boldsymbol{\rho}_p) g_{i}(\boldsymbol{\rho}_t)|_{i=\sigma,\omega,\rho}]$ account for the meson exchange between the nucleons of the projectile and target nuclei \cite{bhuy22}. The results of relativistic R3Y and DDR3Y NN potentials are also compared with the non-relativistic M3Y (Michigan 3 Yukawa) NN potential. As the name suggests, the M3Y NN potential contains three Yukawa terms. Here, we have used both the Paris (denoted here as PM3Y) and Reid (denoted as RM3Y) versions of the M3Y NN potentials, which are written as, 

\begin{eqnarray}
V_{eff}^{PM3Y}(r)=11061.6 \frac{e^{-4r}}{4r}-2537.5\frac{e^{-2.5r}}{2.5r}+J_{00}\delta(r). 
\label{m3y}
\end{eqnarray}\begin{eqnarray}
V_{eff}^{RM3Y}(r)=7999 \frac{e^{-4r}}{4r}-2140\frac{e^{-2.5r}}{2.5r}+J_{00}\delta(r). 
\label{m3y}
\end{eqnarray}
Here, the originally proposed values of OPEP i.e.  $J_{00}=-592$ MeV fm$^{3}$ for PM3Y and $J_{00}=-276$ MeV fm$^{3}$ for RM3Y are used. The density-independent  M3Y interaction have been observed to fail in saturating cold nuclear matter (NM) \cite{bert77,anan83}. Consequently, density dependence is included in the original M3Y interaction via multiplying it by a density-dependent factor $F(\boldsymbol{\rho})$, i.e, 
\begin{eqnarray}
V_{eff}^{M3Y}(\boldsymbol{\rho},r)= F(\boldsymbol{\rho})V_{eff}^{M3Y}(r). 
\label{ddm3y}
\end{eqnarray}
Here, \textbf{$\rho$} denotes the density at the midpoint of the nucleon separation and is obtained within the frozen density approximation (FDA) \cite{satc79,kobo82,khoa94} to calculate the density-dependent M3Y (DDM3Y) NN potentials. Numerous versions of the density-dependent $F(\boldsymbol{\rho})$ factor are available in literature, which are fitted to match the saturation properties of the cold NM. In the present study, we have used the BDM3Y1 ($K=232$ MeV)  version \cite{khoa95} of the Reid density-dependent  M3Y (RDDM3Y) NN interaction and CDM3Y5 ($K=241$ MeV) version \cite{khoa97} of the Paris density-dependent  M3Y (PDDM3Y) NN interaction as they provide the nuclear matter incompressibility (K) value within its present acceptable range.  
The quadrupole deformation parameter ($\beta_2$) for the deformed target nuclei under consideration is determined using the RMF formalism on an axially deformed basis. Additionally, the spherically symmetric radial densities for both projectile and target nuclei are computed using the  NL3$^*$ \cite{lala09} and DDME2 \cite{lala05} parameter sets. For the considered reactions involving deformed targets, the impact of nuclear shape degrees of freedom and orientation is further included through the radius vector in the spherical symmetric target densities. The nuclear radius of an axially deformed nucleus can be written in terms of spherical harmonic expansion as \cite{moll16, bohr52,bohr53}, 
\begin{eqnarray}
{r}_t(\beta_2,\theta_2)={r}_{0t}[1+\sqrt{(5/4\pi)}\beta_2 P_2(cos\theta_2)].
\label{drad}
\end{eqnarray}
Here, $r_{0t}$ is the corresponding spherically symmetric radius. The RMF densities along with the R3Y, DDR3Y, RM3Y, PM3Y, RDDM3Y and PDDM3Y NN interactions are used to calculate the nuclear interaction potential using Eq. \eqref{fold}. The interplay of attractive nuclear potential and repulsive Coulomb and centrifugal potentials results in the formation of the fusion barrier. The properties of this fusion barrier, i.e., barrier height ($V_B^\ell$), barrier position ($R_B^\ell$) and barrier curvature ($\hbar\omega_\ell$) are further used to evaluate the nuclear fusion probability. Several analytical expressions have been formulated to compute the barrier transmission coefficient to avoid its tedious numerical evaluation \cite{jiang21,toub17,canto20,hill53}. The Hill-Wheeler formula obtained under the parabolic barrier approximation \cite{hill53} is one of the well-known approaches that has been used frequently to obtain the barrier transmission coefficient for the heavy-ion fusion at around and above the barrier energies, especially for reactions involving intermediate and heavy-mass nuclei \cite{jiang21,canto20,bhuy18,rana22,bhuy22}. As the reactions considered in the present study also involve the nuclei from the same mass regions, the Hill-Wheeler approximation of the parabolic barrier is used here to obtain the penetrability ($P_{\ell}$) as 
 \begin{eqnarray}
P_\ell(E_{c.m},\theta_2)=\Bigg[1+exp\bigg(\frac{2 \pi (V_{B}^{\ell}(\theta_2)-E_{c.m.})}{\hbar \omega_{\ell}(\theta_2)}\bigg)\Bigg]^{-1}. 
\end{eqnarray}

Here, $E_{c.m.}$ denotes the energy of the target-projectile system in the centre of the mass frame. Finally, the fusion and/or capture cross-section is
calculated using the $\ell-$summed Wong model \cite{bhuy18,kuma09,wong73}. In the $\ell-$summed Wong model, the actual angular momentum dependence of interaction potential is taken into account, and the cross-section is written in terms of $\ell-$partial waves \cite{bhuy18,kuma09,wong73} as, 
\begin{eqnarray}
\sigma(E_{c.m.},\theta_2)=\frac{\pi}{k^{2}} \sum_{\ell=0}^{\ell_{max}}(2\ell+1)P_\ell(E_{c.m},\theta_2).
\label{crs}
\end{eqnarray}
Here, $k=\sqrt{\frac{2 \mu E_{c.m.}}{\hbar^{2}}}$. The $\ell_{max}$ values are calculated using the sharp cut-off model \cite{beck81} at the above barrier energies. These $\ell_{max}$-values are equivalent to the critical angular momenta (which is also symbolized as $\ell_c$ or $\ell_{cr}$) for complete fusion \cite{beck81}. Since the sharp cut-off model is only valid for energies above the barrier, an energy-dependent extrapolation is employed to estimate the $\ell_{max}$-values at sub-barrier energies. The $\ell-$summed cross-section is obtained using Eq. (\ref{crs}) at different target orientation angles ($\theta_2 = 0$ to $\pi/2$). Additionally, as the symmetry axis of the deformed target nucleus is not fixed at a particular angle during the nuclear collision, the cross-section is integrated over the target orientation angle  ($\theta_2$). This approach is widely used to obtain the total cross-section in numerous nuclear fusion studies \cite{kuma09,long08,lari16,rash96,arit12,sun23}. For the reactions involving a spherical projectile and a deformed target, the total integrated cross-section can be expressed as, 
\begin{eqnarray}
\sigma_{int}(E_{c.m.})=\int_{0}^{\pi/2} \sigma(E_{c.m.},\theta_2) sin\theta_2 d\theta_2.
\label{icrs}
\end{eqnarray}

The theoretical formalism described above is used for the systematic and quantitative analysis of nuclear fusion for  $^{16}$O+$^{144}$Sm, $^{48}$Ca+$^{208}$Pb, $^{16}$O+$^{154}$Sm and $^{48}$Ca+$^{238}$U reactions. The calculated results are described in detail in the next section. 

\begin{figure*}
\centering
\includegraphics[scale=0.3]{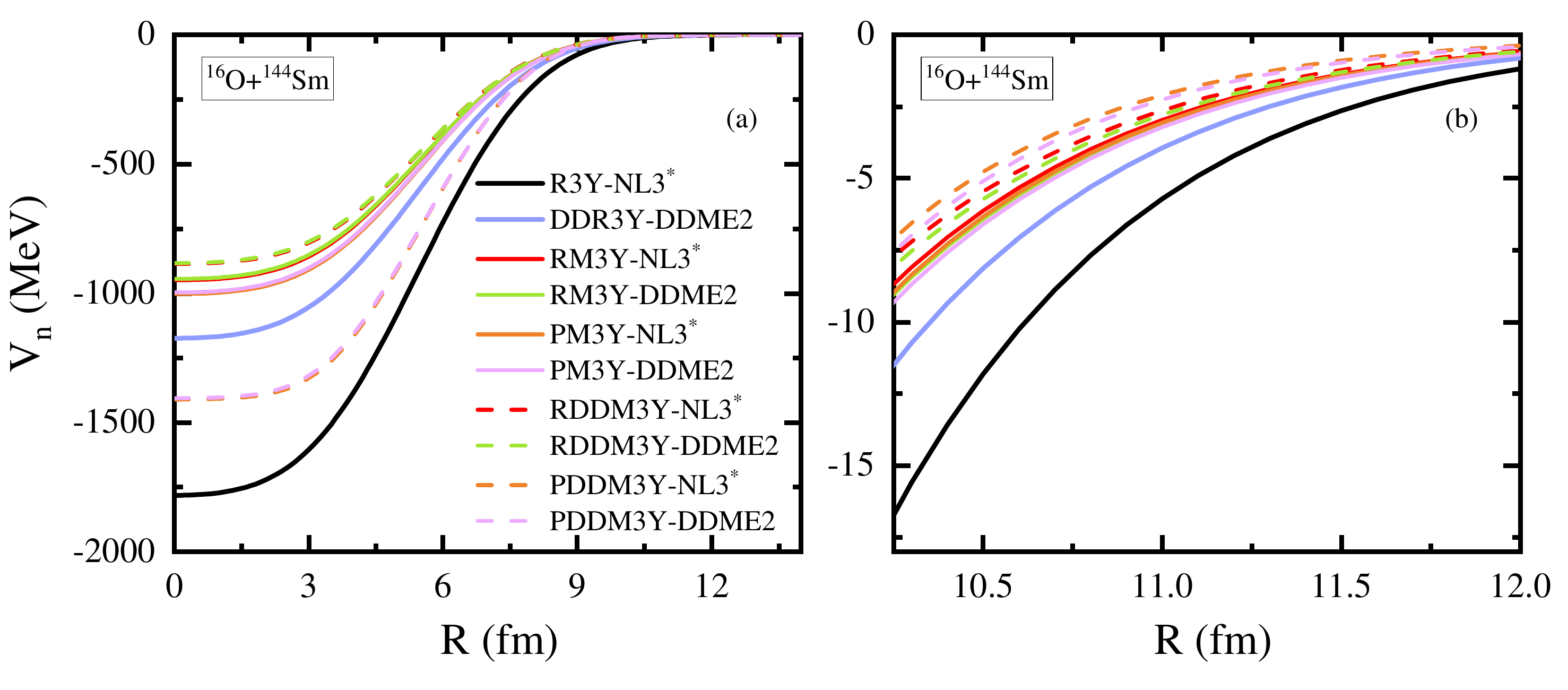}
\caption{(a) The nuclear potential $V_n$ (MeV) as a function of inter-nuclear separation R (fm) calculated using different NN interactions and densities for the $^{16}$O+$^{144}$Sm reaction. (b) magnified view of the surface interaction region of (a).}
\label{fig1}
\end{figure*}
\section{RESULTS AND DISCUSSION}
\label{results}
Nuclear fusion of two heavy-ions can be affected by many entrance channel properties such as the energy, angular momentum, isospin, mass and charge asymmetry, binding energy, nuclear shell closure, pairing energy, nuclear shape and orientations etc. The impact of these factors is usually included in the description of nuclear fusion through the interaction potential formed between two colliding heavy ions. As accounting for all the entrance channel effects in the description of heavy-ion fusion at once becomes theoretically tedious, stepwise refinements are done in the theoretical models to systematically understand the role of one factor or another. Here, we aim to quantitatively analyse a few entrance channel effects on the nuclear potential obtained within the double-folding approach. The density distributions of the interacting nuclei and an effective nucleon-nucleon (NN) interaction are the primary inputs for the computation of nuclear interaction potential within the double-folding approach. Here, we have considered six different effective NN interactions to study the heavy fusion dynamics. These NN interactions include the relativistic effective interaction (R3Y) potential for the NL3$^*$ parameter set and the density-dependent R3Y (DDR3Y) NN potential for the DDME2 parameter set. The results of these relativistic effective interactions are also compared with the non-relativistic Reid and Paris versions of Michigan 3 Yukawa  (M3Y) NN interaction and also their density-dependent versions, namely BDM3Y1  and CDM3Y5 NN interactions.  It is to be noted that Reid M3Y NN interaction is denoted as RM3Y, and Paris M3Y NN interaction is denoted as PM3Y from here onward. Also, the  Reid density-dependent M3Y effective NN interaction is denoted as RDDM3Y and Paris density-dependent M3Y effective NN interaction is denoted as PDDM3Y.  The densities of the fusing nuclei are obtained from the relativistic mean-field (RMF) approach for the NL3$^*$ set and the Relativistic-Hartree-Bogoliubov (RHB) formalism for the DDME2 set. Folding these NN potentials and densities results in 10 different nuclear potentials, which are named R3Y-NL3$^*$, DDR3Y-DDME2, RM3Y-NL3$^*$, RM3Y-DDME2, PM3Y-NL3$^*$, PM3Y-DDME2, RDDM3Y-NL3$^*$, RDDM3Y-DDME2, PDDM3Y-NL3$^*$ and PDDM3Y-DDME2. Here, R3Y-NL3$^*$ represents that the nuclear potential is obtained by folding the R3Y NN potential with the NL3$^*$ densities. A similar notation will be used from here onward.
\begin{figure*}
\centering
\includegraphics[scale=0.35]{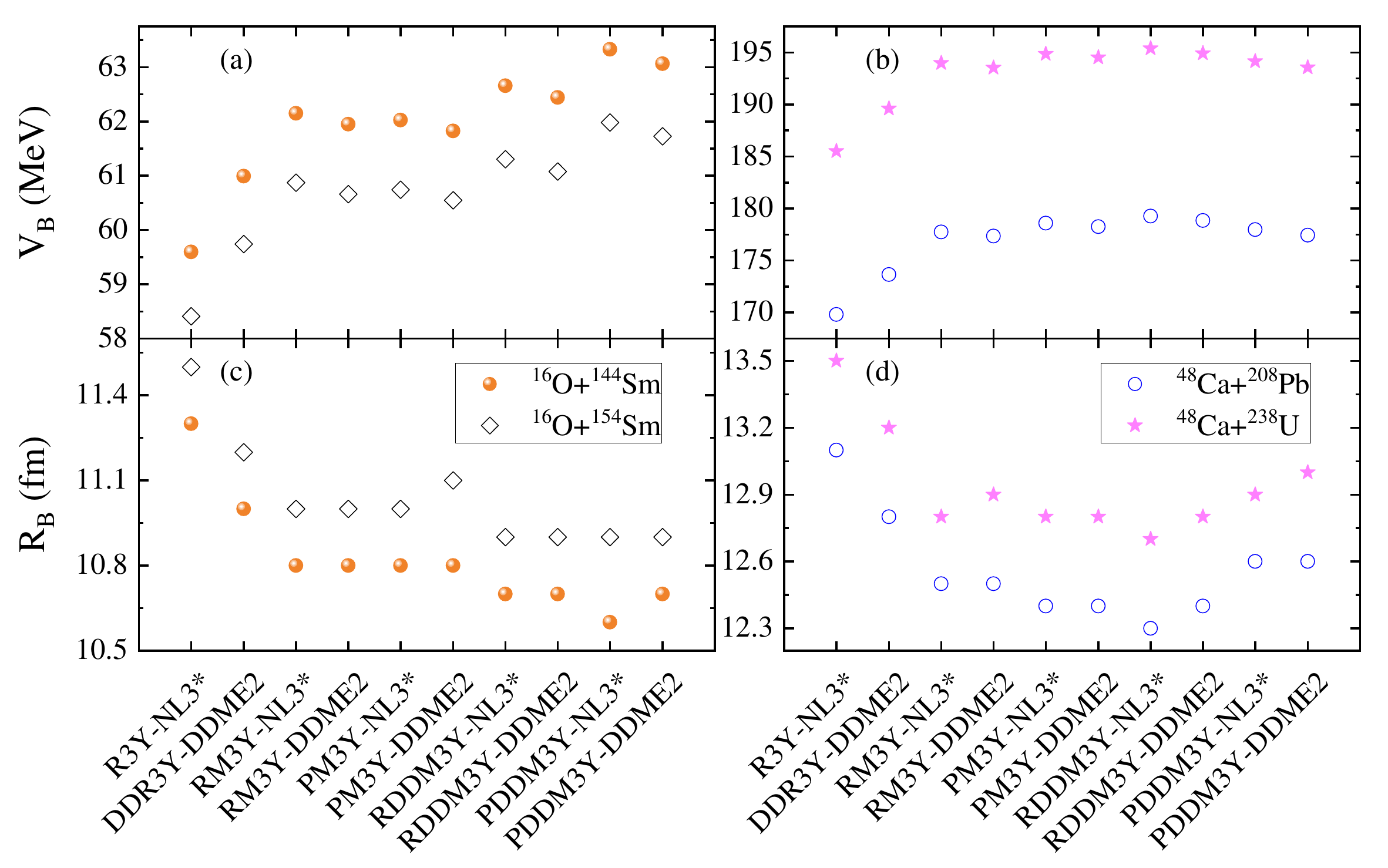}
\caption{The heights $V_B$ (MeV) (upper panel) and positions $R_B$ (fm) (lower panel) of the s-wave ($\ell=0$ $\hbar$) fusion barriers obtained using different nuclear potentials.}
\label{fig2}
\end{figure*}

Figure \ref{fig1} (a) shows the different nuclear potentials considered in the present study as a function of internuclear separation R (fm) for the illustrative case of $^{16}$O+$^{144}$Sm reaction. The magnified view of the nuclear potential at the larger $R$, i.e. the surface interaction region, which plays the most significant role in nuclear fusion, is shown in Fig. \ref{fig1} (b). The relativistic R3Y NN potential and densities for the NL3$^*$ parameter set are observed to provide the most attractive nuclear potential for the $^{16}$O+$^{144}$Sm reaction. The medium-dependent DDR3Y NN potential obtained within the RHB approach for DDME2 is found to give a less attractive nuclear potential compared to the R3Y NN potential. This is because the introduction of in-medium effects in terms of density-dependent nucleon-meson couplings within the RHB formalism in the description of NN interaction potential results in a comparatively repulsive nuclear potential. On comparing the results of relativistic NN interactions with the non-relativistic RM3Y and PM3Y NN potentials, the relativistic R3Y and DDR3Y potentials are noted to provide a more attractive nuclear potential.  Moreover, the Paris version of the density-independent M3Y NN interaction gives a slightly more attractive nuclear potential than the Reid version. Furthermore, the inclusion of density dependence in the PM3Y NN potential results in a more attractive nuclear potential at smaller R, and this trend is inverted at the larger R. On the other hand, the inclusion of density-dependence in the RM3Y NN potential gives a repulsive nuclear potential at all R compared to the density-independent RM3Y NN potential.

A comparison of DDM3Y and DDR3Y NN potentials shows that the DDR3Y NN potential provides a more attractive nuclear potential at the surface interaction region (larger R) significant for nuclear fusion. It is worth noting here that in DDR3Y NN potential, the in-medium effects are included in terms of density-dependent nucleon-meson couplings (see Eq. \ref{r3y}) within the RHB formalism. However, in-medium effects are included in the DDM3Y NN potential through density-dependent factors (see Eq. \ref{ddm3y}). Moreover, different forms of this density-dependent factor, i.e., CDM3Y5 version \cite{khoa97} of PDDM3Y and BDM3Y1  version \cite{khoa95} of the RDDM3Y NN interaction, are used here to include the density-dependence in Paris and Reid version of M3Y potential. The frozen density approximation \cite{khoa97, khoa95} is used in the DDM3Y NN potential. However, for the case of DDR3Y NN potential, the in-medium effects are included directly in terms of both projectile and target density-dependent nucleon-meson couplings to account for the meson exchange between the nucleons of the projectile and target nuclei. On the other hand, the non-linear couplings of $\sigma-$meson (terms with $g_2$ and $g_3$) in the relativistic R3Y NN potential obtained here for NL3$^*$ parameter set account effectively for the in-medium effects. All these underlying distinctions in the description of effective NN interactions result in different forms of nuclear potentials, which can be seen in Fig. \ref{fig1}. When comparing the nuclear potential obtained for the densities of NL3$^*$ and DDME2, it is observed that the densities of DDME2 provide a slightly more attractive nuclear potential.

For a more systematic investigation of the impact of the different NN interactions and densities discussed above on heavy-ion fusion, next we have compared the fusion barrier characteristics for four reactions namely, $^{16}$O+$^{144}$Sm, $^{48}$Ca+$^{208}$Pb, $^{16}$O+$^{154}$Sm and $^{48}$Ca+$^{238}$U. As discussed previously, the interplay of the attractive nuclear and repulsive Coulomb and centrifugal potentials results in the formation of the fusion barrier. Figure \ref{fig2} shows the heights $V_B$ (MeV) (upper panel) and positions $R_B$ (fm) (lower panel) of the s-wave ($\ell=0$ $\hbar$) fusion barriers obtained using different nuclear potentials. The left panel of the Fig. \ref{fig2} shows the barrier characteristics for the  $^{16}$O+$^{144}$Sm (orange spheres) and $^{16}$O+$^{154}$Sm (black squares) reactions. The barrier characteristics for the  $^{48}$Ca+$^{208}$Pb (blue circles) and $^{48}$Ca+$^{238}$U (magenta stars) are shown in the right panel of Fig. \ref{fig2}. It can be noted from the upper panel of Fig. \ref{fig2} that the relativistic R3Y NN potential and densities obtained for the NL3$^*$ parameter set give the lowest fusion barrier for all the considered reactions. The non-relativistic M3Y NN interactions are noted to give a higher fusion barrier than the relativistic NN interactions for all the considered reactions. Moreover, the inclusion of in-medium effects is observed to increase the barrier height for both the R3Y and M3Y NN interactions. This increment is relatively higher for the relativistic DDR3Y NN potential as compared to the DDM3Y NN potential. The trend of the barrier positions is noted to be opposite to that of the barrier heights. This means that the higher fusion barrier is formed at a smaller interaction radius (smaller R) and vice-versa. Moreover, the highest barrier is noted for the  PDDM3Y-NL3$^*$, i.e., for nuclear potential obtained using density-dependent Paris M3Y NN potential and NL3$^*$ densities for the  $^{16}$O+$^{144,154}$Sm reactions. On the other hand, RDDM3Y-NL3$^*$ nuclear potential is noted to give the highest barriers for the $^{48}$Ca+$^{208}$Pb and $^{48}$Ca+$^{238}$U reactions. The nuclear densities obtained for the DDME2 parameter set are observed to give lower barrier heights as compared to the NL3$^*$ densities for all the considered reactions. In addition, the difference in the barrier characteristics obtained for different nuclear potentials is more pronounced for $^{48}$Ca+$^{208}$Pb and $^{48}$Ca+$^{238}$U reactions, which lead to the formation of compound nuclei in the superheavy region.  All of these observations infer that the characteristics of the fusion barrier are sensitive to various factors that are included in the descriptions of the interaction potential formed between two fusion heavy ions.
\begin{figure*}
\centering
\includegraphics[scale=0.17]{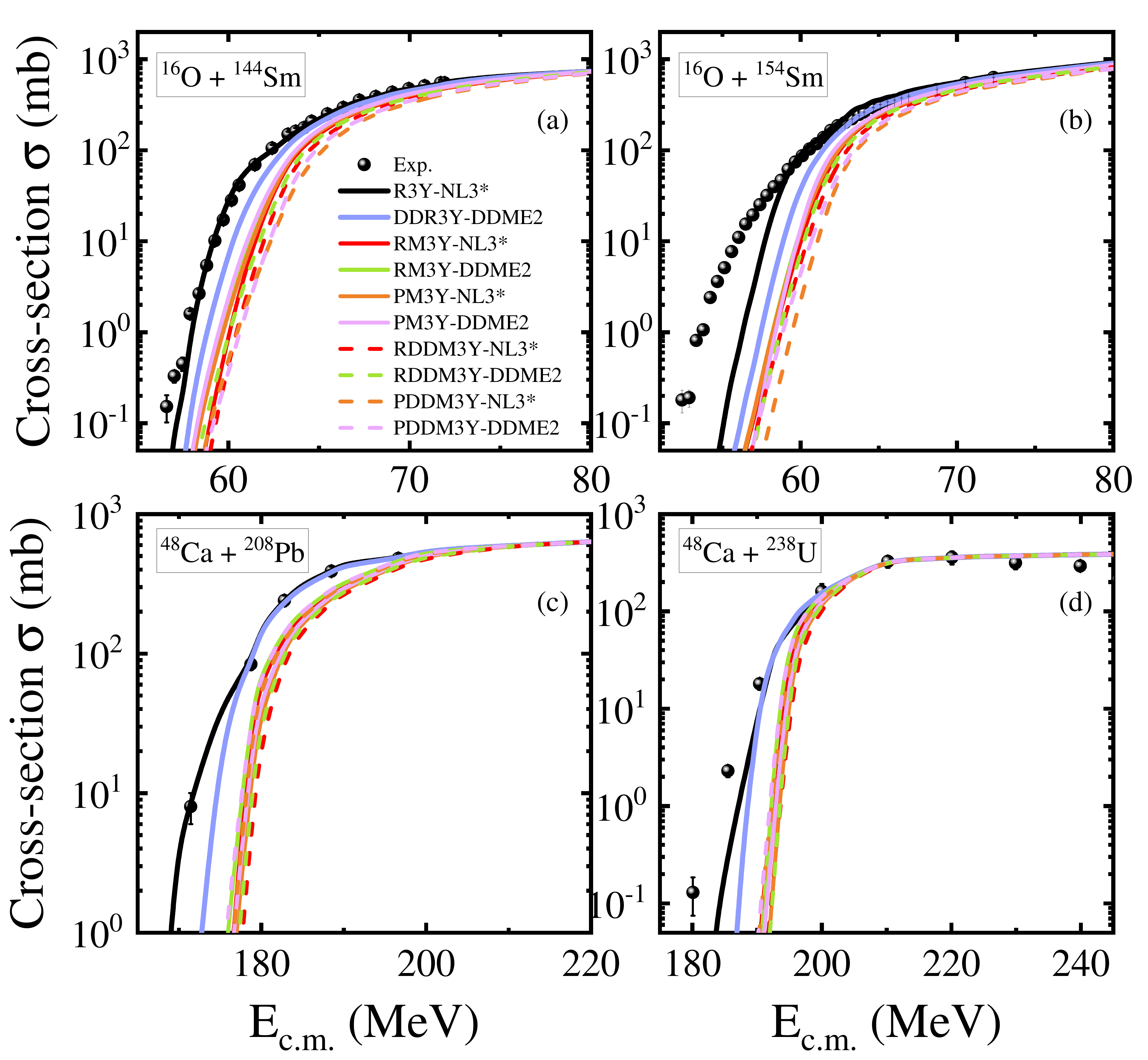}
\caption{ The cross-section $\sigma$ (mb) calculated using $\ell-$summed Wong model supplemented with different nuclear potentials for (a) $^{16}$O+$^{144}$Sm, (b) $^{16}$O+$^{154}$Sm,  (c) $^{48}$Ca+$^{208}$Pb and (d) $^{48}$Ca+$^{238}$U reactions. The experimental data from \cite{morton94,prok08,leigh95,nishi12} is also shown for comparison. }
\label{fig3}
\end{figure*}

To assess the validity of various nuclear potentials to study the nuclear fusion mechanism, next, we have calculated the fusion and/or capture cross-section for all the considered reactions. The characteristics of the fusion barrier are used to calculate the probability of barrier penetration within the Hill-Wheeler approach \cite{hill53}. In this approach, the fusion barrier is assumed to have a parabolic shape to simplify the computation of the barrier transmission coefficient \cite{hill53}. This approach has been observed to provide reliable results for the nuclear fusion of various heavy-ion reactions \cite{jiang21,canto20,bhuy18,rana22,bhuy22}. Next, the fusion and/or capture cross-section is calculated using the $\ell-$summed Wong model \cite{bhuy18,kuma09,wong73} with $\ell_{max}$ values obtained from the sharp cut-off model \cite{beck81} model at the above-barrier energies. Further, energy-dependent extrapolation is used to obtain the $\ell_{max}$ values at sub-barrier energies. Figure \ref{fig3} shows the cross-section $\sigma$ (mb) calculated using $\ell-$summed Wong model supplemented with different nuclear potentials for (a) $^{16}$O+$^{144}$Sm, (b) $^{16}$O+$^{154}$Sm,  (c) $^{48}$Ca+$^{208}$Pb and (d) $^{48}$Ca+$^{238}$U reactions. The microscopic nuclear potential calculated using relativistic R3Y NN interaction and nuclear densities for the NL3$^*$ parameter set is noticed to give the highest cross-section for all the considered reactions at below and around the barrier centre of mass energies ($E_{c.m.}$). The medium-dependent relativistic DDR3Y NN potential obtained within the RHB approach for the DDME2 parameter set is observed to provide a lower cross-section than the relativistic R3Y NN potential obtained for non-linear NL3$^*$ parametrization. Moreover, the density-dependent versions of Paris and Reid M3Y NN interactions are also noted to give a lower cross-section than their medium-independent versions. This infers that the inclusion of in-medium effects in the description of NN interactions leads to the hindrance in nuclear fusion.

Further, on comparing the results of relativistic and non-relativistic NN interactions, it is noted that the relativistic effective NN interactions provide comparatively higher cross-sections for all the reactions under study. This observation persists for both the Reid as well as Paris versions of the non-relativistic M3Y NN interactions. It is to be noted here that we have adopted the originally proposed values of one-pion exchange potential (OPEP), i.e.  $J_{00}=-592$ MeV fm$^{3}$ for PM3Y and $J_{00}=-276$ MeV fm$^{3}$ for RM3Y. Also, the same value of $\ell_{max}$ is maintained at a given  $E_{c.m.}$ for all the nuclear potentials. Comparison of cross-section obtained using nuclear densities obtained within RMF formalism for the non-linear NL3$^*$ set and RHB approach for the density-dependent DDME2 set shows that the DDME2 densities give slightly enhanced cross-section for all the considered reactions. Moreover, the cross-section obtained using different nuclear potentials is noted to overlap at the above-barrier energies due to the suppression of nuclear structure effects and the dominance of angular momentum effects in this energy region. To assess the validity of different nuclear potentials to describe the nuclear fusion mechanism, the calculated cross-sections are also compared with the experimental data from \cite{morton94,prok08,leigh95,nishi12}. The R3Y-NL3$^*$ nuclear potential is observed to provide a better match to the experimental data than the other nuclear potentials considered in the present analysis. Both the Paris and Reid versions of the non-relativistic M3Y NN potentials folded with both NL3$^*$ and DDME2 densities are noticed to underestimate the fusion and/or capture cross-section at below and around the barrier energies for all the reactions under study. The inclusion of in-medium effects in both relativistic and non-relativistic effective NN interactions results in a relatively repulsive nuclear potential, which further leads to the underestimation of the cross-section.

\begin{figure*}
\centering
\includegraphics[scale=0.12]{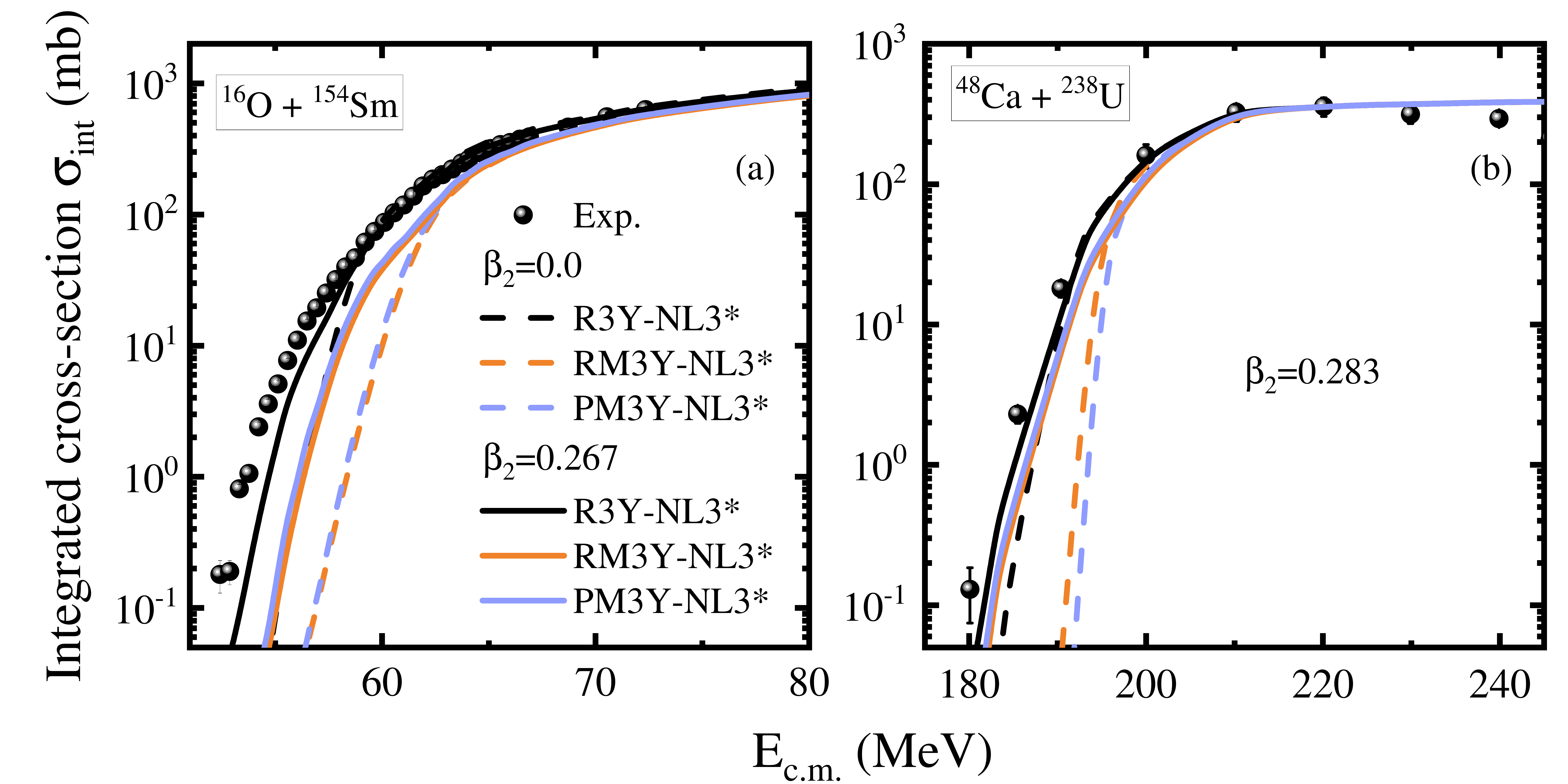}
\caption{The total integrated cross-section $\sigma_{int}$ (mb) obtained using the R3Y (black lines), RM3Y (orange lines) and PM3Y (blue lines)  NN potentials folded with spherical (dashed lines) and deformed (solid lines) NL3$^*$ densities for (a) $^{16}$O+$^{154}$Sm and (b) $^{48}$Ca+$^{238}$U reactions. The experimental data \cite{leigh95,nishi12} is also given for comparison.}
\label{fig4}
\end{figure*}

It is worth noting here that all the cross-sections plotted in Fig. \ref{fig3} are calculated assuming the spherical symmetry of the fusing projectile and target nuclei. It can be noted clearly from Fig. \ref{fig3} that the calculated fusion and/or capture cross-section using relativistic R3Y NN potential provides a better match to the experimental data as compared to that obtained using the DDR3Y, RM3Y and PM3Y NN interactions for both type reactions considered in the present analysis, i.e., $^{16}$O+$^{144}$Sm and $^{48}$Ca+$^{208}$Pb reactions containing the spherical reactions partners as well as $^{16}$O+$^{154}$Sm and $^{48}$Ca+$^{238}$U reactions involving deformed targets. However, the R3Y-NL3$^*$ nuclear potential provides much better match to the experimental data for $^{16}$O+$^{144}$Sm and $^{48}$Ca+$^{208}$Pb reactions as compared to the $^{16}$O+$^{154}$Sm and $^{48}$Ca+$^{238}$U reactions. This is because both $^{154}$Sm and  $^{238}$U nuclei are observed to be deformed in their ground state \cite{raman01,prit12}, and the cross-sections shown in Fig. \ref{fig3} are obtained assuming their spherical shape. As discussed previously, the nuclear fusion mechanism is observed to be affected by various structural properties of interacting projectile and target nuclei and step-wise refinements and improvements are requisite to account for these factors in the description of nuclear interaction potential.

Following these, here we have also moved a step ahead in our calculation of microscopic nuclear potential within the RMF formalism. The impact of nuclear shape degrees of freedom is included in the computation of nuclear potential in terms of the deformed densities of the target nuclei (see Eq. \eqref{fold}). The impact of nuclear deformations is accounted in the target density distributions through the nuclear radius using Eq. \eqref{drad}. As discussed above, the inclusion of density-dependence in both R3Y and M3Y NN potentials leads to the underestimation of fusion and/or capture cross-section. Following this, we have done the calculations with the inclusion of nuclear shape degrees of freedom for the density-independent versions relativistic R3Y and non-relativistic RM3Y and PM3Y  NN potentials. These NN potentials are folded with the deformed nuclear densities obtained for the non-linear NL3$^*$ parameter set. The quadrupole deformations ($\beta_2$) for the $^{154}$Sm and  $^{238}$U nuclei are also obtained from the axially deformed RMF theory with the NL3$^*$ parameter set \cite{lala09}. When we consider the impact of target deformations, the interaction potential also becomes $\theta_2$-dependent (see Eq. \eqref{vtot}), with $\theta_2$ being the orientation angle between the symmetry axis of the axially deformed target nucleus and inter-nuclear separation ($R$). Here, the interaction potential and the barrier characteristics are obtained at each target orientation angle  ($\theta_2$). A systematic analysis of orientation dependence of the fusion barrier characteristics can be found in our recent study \cite{rana23}. However, as the target nuclei can not be oriented at a specific angle during the nuclear collision, the cross-section is integrated over the target orientation angle  ($\theta_2$) as per Eq. \eqref{icrs}.

\begin{figure}
\centering
\includegraphics[scale=0.08]{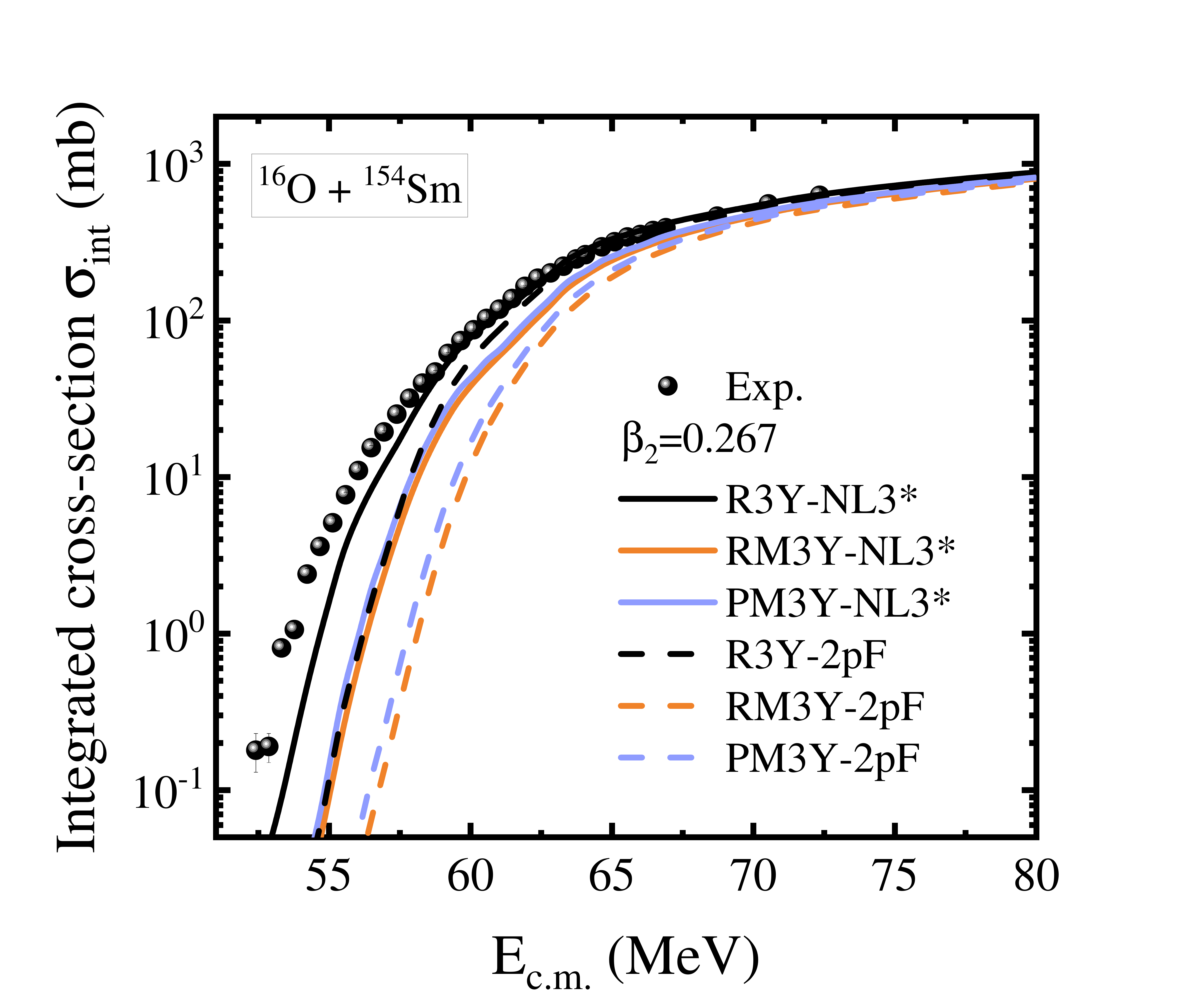}
\caption{The total integrated cross-section $\sigma_{int}$ (mb) obtained for the $^{16}$O+$^{154}$Sm reaction using the R3Y (black lines), RM3Y (orange lines) and PM3Y (blue lines)  NN potentials folded with nuclear densities obtained from 2pF (dashed lines) and RMF-NL3$^*$ (solid lines) approaches. The experimental data \cite{leigh95} is also given for comparison.}
\label{fig5}
\end{figure}

Figure \ref{fig4} shows the total integrated cross-section $\sigma_{int}$ (mb) obtained using the R3Y (black lines), RM3Y (orange lines) and PM3Y (blue lines)  NN potentials folded with spherical (dashed lines) and deformed (solid lines) NL3$^*$ densities for (a) $^{16}$O+$^{154}$Sm and (b) $^{48}$Ca+$^{238}$U reactions. The values of the quadrupole deformations  ($\beta_2$) for the $^{154}$Sm and  $^{238}$U nuclei from the axially deformed RMF-NL3$^*$ formalism are also mentioned in the respective panels of Fig. \ref{fig4}. It can be noted from here that taking into account the impact of target quadrupole deformations in the computation of nuclear interaction potential results in the enhancement of the fusion and/or capture cross-section at the below-barrier energies. This infers that deviation of the shape of the target nucleus from spherical ($\beta_2=0.0$) to prolate ($\beta_2>0.0$) increases the probability of nuclear fusion.  Further, the inclusion of nuclear shape degrees of freedom in the calculations of the nuclear interaction potential also results in a better match to the experimental data \cite{leigh95,nishi12} for reactions involving deformed target nuclei. On comparing the results obtained using different NN potentials, it is observed that the relativistic R3Y NN potential provides a much more reasonable match to experimental data as compared to both the Paris and Reid versions of M3Y NN potential for both the reactions under study. All these observations infer that the relativistic R3Y effective NN potential and RMF nuclear densities for the non-linear  NL3$^*$ parameter set provide a more reliable description of the heavy-ion fusion as compared to both the Paris and Reid versions of the non-relativistic M3Y NN potential for all the considered reactions. These observations also resolve the concerns raised in Ref. \cite{chus22} regarding the use of spherical densities of the target nuclei in \cite{bhuy20}, since here the nuclear potential obtained within the double folding approach supplemented with R3Y NN potential and nuclear densities from the RMF formalism is observed to provide a reasonable description of heavy-ion fusion for reactions containing spherical as well as deformed target nuclei.
\begin{figure*}
\centering
\includegraphics[scale=0.5]{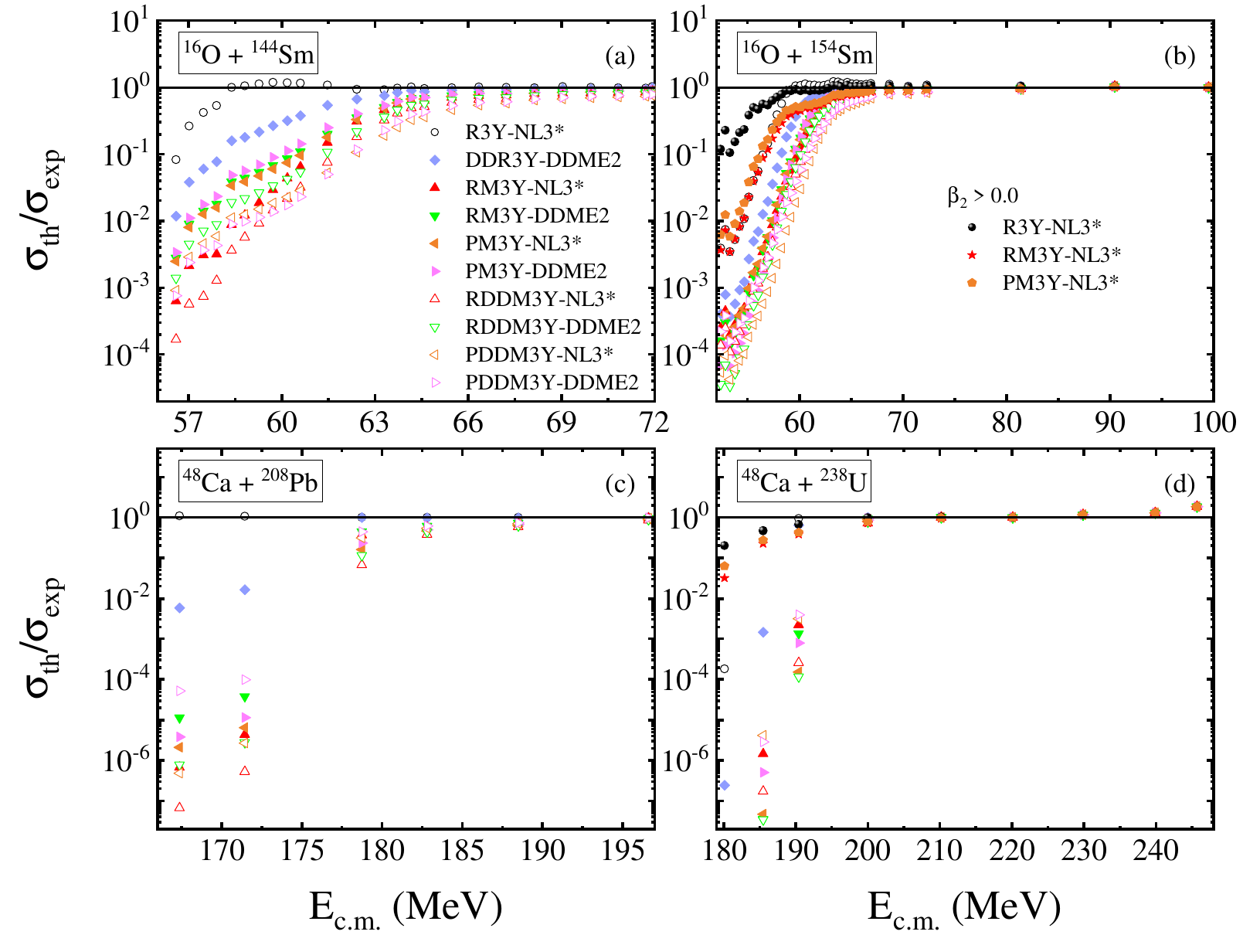}
\caption{ The ratio of calculated theoretical cross-section ($\sigma_{th}$) to the experimental data ($\sigma_{exp}$) for (a) $^{16}$O+$^{144}$Sm, (b) $^{16}$O+$^{154}$Sm,  (c) $^{48}$Ca+$^{208}$Pb and (d) $^{48}$Ca+$^{238}$U reactions.}
\label{fig6}
\end{figure*}

Furthermore, the nuclear potentials obtained by folding the Paris M3Y NN potential with the deformed densities obtained from the two-parameter Fermi (2pF) formula were observed to provide a reasonable agreement with the experimental data for $^{48}$Ca+$^{238}$U reaction \cite{chus22}. These interesting observations entice us to compare the results of 2pF densities with those obtained within the self-consistent RMF formalism. For this, we have compared the cross-section calculated using the quadrupole deformed 2pF and RMF-NL3$^*$ densities for the illustrative case of   $^{16}$O+$^{154}$Sm since the experimental cross-section is available for this reaction at far sub-barrier energies with much smaller step-sizes. Here, we have followed the methodology of Ref. \cite{gont06} to obtain the 2pF densities for spherical $^{16}$O projectile and quadrupole deformed $^{154}$Sm target. It is to be noted here that the $\beta_2=0.267$ from RMF-NL3$^*$ formalism is used for both the 2pF and RMF densities, and the impact of higher order deformations of the $^{154}$Sm is not taken into account. Figure \ref{fig5} shows the total integrated cross-section $\sigma_{int}$ (mb) obtained for the $^{16}$O+$^{154}$Sm reaction using the R3Y (black lines), RM3Y (orange lines) and PM3Y (blue lines)  NN potentials folded with nuclear densities obtained from 2pF (dashed lines) and RMF-NL3$^*$ (solid lines) approaches. The nuclear densities obtained from the self-consistent RMF formalism for the well-known NL3$^*$ set are observed to provide a higher sub-barrier cross-section as compared to the 2pF densities. Moreover, the RMF-NL3$^*$ densities are noted to provide more reasonable agreement to the experimental data than the 2pF densities for $^{16}$O+$^{154}$Sm reaction. The Paris M3Y (PM3Y) NN interactions are observed to provide a slightly higher cross-section and a better match to experimental data as compared to the Reid version but it still underestimates the experimental cross-section at the sub-barrier energy region. However, these observations might vary for other reactions since the results of 2pF densities depend upon the choice of radius and surface diffuseness parameters. This gives scope for a more comprehensive study with various heavy-ion reactions containing the nuclei from different regions of the nuclear chart to explore the impact of different density distributions on the nuclear fusion mechanism.

For a quantitative comparison of the theoretical cross-sections ($\sigma_{th}$) calculated using different nuclear potentials with those obtained experimentally ($\sigma_{exp}$), the ratio $\sigma_{th}/ \sigma_{exp}$ is plotted as a function of centre of mass energies ($E_{c.m.}$) in Fig. \ref{fig6} for (a) $^{16}$O+$^{144}$Sm, (b) $^{16}$O+$^{154}$Sm,  (c) $^{48}$Ca+$^{208}$Pb and (d) $^{48}$Ca+$^{238}$U reactions under study. The R3Y NN potential folded with RMF-NL3$^*$ densities is noted to provide the least deviation from the experimental data for $^{16}$O+$^{144}$Sm and $^{48}$Ca+$^{208}$Pb reactions containing the spherical reactions partners. This observation also persists for $^{16}$O+$^{154}$Sm and $^{48}$Ca+$^{238}$U reactions after the inclusion of target quadrupole deformations. Moreover, both the Paris as well as Reid versions of non-relativistic  M3Y NN interaction are noted to provide more deviation from the experimental data as compared to the relativistic R3Y and DDR3Y NN interactions. Further, the density-dependent versions of R3Y, RM3Y and PM3Y NN potentials are noticed to give higher deviation from the experimental data as compared to their density-independent versions. A more careful inspection of Fig. \ref{fig6} shows that however, the R3Y-NL3$^*$ nuclear potential provide better match than the other considered nuclear potential, it still underestimates the cross-section for $^{16}$O+$^{144,154}$Sm and $^{48}$Ca+$^{238}$U at deep sub-barrier energies. This mismatch can be due to many underlying nuclear structure effects which are not taken into account in the present study. For example, the underestimation noted for  $^{16}$O+$^{154}$Sm and $^{48}$Ca+$^{238}$U reactions can be attributed to the higher order deformations of $^{154}$Sm and  $^{238}$U nuclei as here we have considered only their quadrupole deformations. All the observations of the present analysis lead to the conclusion that the dynamics of nuclear fusion are affected by various structural properties of the interacting nuclei. Thus, meticulous and systematic refinements in the various theoretical models are necessary for a more comprehensive understanding of the heavy-ion fusion mechanism.

\section{Summary and Conclusions} 
\label{summary}
The heavy-ion fusion mechanism is explored using different nuclear potentials obtained within the double-folding approach. In this study, we employ various effective nucleon-nucleon (NN) interactions and densities to calculate the nuclear potential within the double-folding framework. The NN interactions include the relativistic effective interaction (R3Y) NN potential, derived from the self-consistent relativistic mean-field (RMF) formalism using the well-adopted NL3$^*$ parameter set, as well as its density-dependent version (denoted as DDR3Y NN potential), which is derived from the Relativistic-Hartree-Bogoliubov (RHB) approach using the density-dependent DDME2 set. The results of these relativistic effective NN interactions are compared with the Reid (denoted as RM3Y) and Paris (denoted as PM3Y) versions of the well-known M3Y (Michigan 3 Yukawa) NN interaction. Additionally, we consider the density-dependent versions of these  M3Y NN interactions, namely BDM3Y1 for Reid (denoted as RDDM3Y) and CDM3Y5  for Paris (denoted as PDDM3Y) NN potentials, in the present analysis. The density distributions of the fusing nuclei are obtained from the RMF approach using the non-linear NL3$^*$ parametrization and from the RHB approach using the DDME2 set.

The underlying distinctions in the descriptions of these NN interactions and densities result in different nuclear potentials. To assess the applicability of these nuclear potentials in studying nuclear fusion, a systematic analysis of fusion barrier characteristics and cross-sections is carried out for the $^{16}$O+$^{144}$Sm and $^{48}$Ca+$^{208}$Pb reactions, which involve spherical reaction partners, as well as for $^{16}$O+$^{154}$Sm and $^{48}$Ca+$^{238}$U reactions, which involve deformed targets. The relativistic R3Y and DDR3Y NN potentials, formulated in terms of nucleon-meson couplings, yield higher cross-sections than the Reid and Paris versions of the non-relativistic M3Y NN potential. Moreover, the inclusion of in-medium effects in both the R3Y and M3Y NN interactions leads to relatively repulsive nuclear potentials, resulting in higher fusion barriers and lower cross-sections for all the considered reactions. However, the reduction in cross-section is more pronounced for the relativistic DDR3Y NN potential, in which in-medium effects are introduced through density-dependent nuclear-meson couplings (see Eq. \eqref{r3y}), than for the non-relativistic RDDM3Y and PDDM3Y NN potentials, where in-medium effects are accounted for by a density-dependent factor (see Eq. \eqref{ddm3y}). Additionally, the NL3$^*$ densities provide slightly higher fusion barriers and lower cross-sections compared to the DDME2 densities.

The fusion and/or capture cross-section, obtained using the $\ell$-summed Wong model supplemented with various nuclear potentials, is also compared with available experimental data for the considered reactions. The R3Y-NL3$^*$ nuclear potential obtained by folding the R3Y NN potential and densities for the NL3$^*$ parameter set provides comparatively better agreement with the experimental data at sub-barrier energies for all the considered reactions. To explore the impact of nuclear structure on fusion dynamics, we next include the effect of target quadrupole deformations in the description of the nuclear interaction potential for the $^{16}$O+$^{154}$Sm and $^{48}$Ca+$^{238}$U reactions, which involve deformed targets. An increase in the total integrated cross-section is noted when the shape of the target nucleus deviates from spherical ($\beta_2 = 0.0$) to prolate ($\beta_2 > 0.0$). This enhancement in the cross-section, due to the inclusion of target quadrupole deformations, also leads to better agreement between the calculated results and the experimental data. Further, the results of density distributions obtained from the self-consistent RMF-NL3$^*$ formalism are compared with those obtained from the two-parameter Fermi (2pF) formula for the illustrative case of $^{16}$O+$^{154}$Sm reactions. The RMF-NL3$^*$ densities are more reliable than the 2pF densities for the considered reaction. Additionally, the ratios of theoretical cross-sections obtained using different nuclear potentials and experimental cross-sections ($\sigma_{th}/\sigma_{exp}$) are analyzed for the $^{16}$O+$^{144}$Sm, $^{48}$Ca+$^{208}$Pb, $^{16}$O+$^{154}$Sm, and $^{48}$Ca+$^{238}$U reactions. From these qualitative and quantitative analyses of fusion and/or capture, cross-sections calculated using different nuclear potentials, the R3Y NN potential and nuclear densities obtained from the RMF formalism for the non-linear NL3$^*$ parameter set are found to provide a reasonable description of heavy-ion fusion for reactions involving both spherical and deformed target nuclei. A more systematic and comprehensive analysis will be carried out in future studies to explore the impact of various nuclear densities, as well as the role of higher-order nuclear deformations, on the fusion mechanism of heavy-ion reactions across different regions of the nuclear chart.

\section*{Acknowledgments}
\noindent
This work is partially supported by the Science and Engineering Research Board (SERB) File No. CRG/2021/001229, and Ramanujan Fellowship File No. RJF/2022/000140.


\end{document}